\documentclass[10pt,a4paper,twocolumn]{article}
\usepackage[utf8x]{inputenc}
\usepackage{amsmath}
\usepackage{multirow}
\usepackage{wrapfig}
\usepackage{bbold}
\usepackage{graphicx}
\usepackage[round]{natbib}
\usepackage{authblk}
\usepackage{fancyhdr}

\begin{document}

\title{Inferring gene expression networks with hubs using a degree weighted Lasso approach}
\author{Nurgazy Sulaimanov\,$^{\text{1,2,}}$\footnote{nurgazy.sulaimanov@bcs.tu-darmstadt.de}, Sunil Kumar\,$^{\text{1,2}}$, Fr\'ed\'eric Burdet\,$^{\text{3}}$, \\ Mark Ibberson\,$^{\text{3}}$,  Marco Pagni\,$^{\text{3}}$,  Heinz Koeppl\,$^{\text{1,2,}}$\footnote{heinz.koeppl@bcs.tu-darmstadt.de}}
\affil{$^{\text{\sf 1}}$Department of Electrical Engineering and Information Technology, $^{\text{\sf 2}}$Department of Biology, Technische Universit\"{a}t Darmstadt, Darmstadt, Germany, $^{\text{\sf 3}}$Swiss Institute of Bioinformatics, Lausanne, Switzerland} 
%\affil{$^{\text{\sf 2}}$Swiss Institute of Bioinformatics, Lausanne, Switzerland}
\affil{This article has been submitted to the Bioinformatics.}

\maketitle

\begin{abstract}
Genome-scale gene networks contain regulatory genes called hubs that have many interaction partners. These genes usually play an essential role in gene regulation and cellular processes. Despite recent advancements in high-throughput technology, inferring gene networks with hub genes from high-dimensional data still remains a challenging problem. Novel statistical network inference methods are needed for efficient and accurate reconstruction of hub networks from high-dimensional data.  \\
To address this challenge we propose DW-Lasso, a degree weighted Lasso (least absolute shrinkage and selection operator) method which infers gene networks with hubs efficiently under the low sample size setting.
Our network reconstruction approach is formulated as a two stage procedure: first, the degree of networks is estimated iteratively, and second, the gene regulatory network is reconstructed using degree information. A useful property of the proposed method is that it naturally favors the accumulation of neighbors around hub genes and thereby helps in accurate modeling of the high-throughput data under the assumption that the underlying network exhibits hub structure.
In a simulation study, we demonstrate good predictive performance of the proposed method in comparison to traditional Lasso type methods in inferring hub and scale-free graphs. We show the effectiveness of our method in an application to microarray data of \textit{E.coli} and RNA sequencing data of Kidney Clear Cell Carcinoma from The Cancer Genome Atlas datasets.
\end{abstract} 

\section{Introduction}
With the advent of high-throughput technologies such as microarrays and RNA sequencing, inference of gene regulatory networks has attracted much scientific interest over the last decade. 
The technologies enable simultaneous measurement of large numbers of genes. This leads to the challenge of inferring large-scale gene regulatory networks from high-dimensional data (\citet{citeulike:10886734}; \cite{hill2016}).
On the other hand, high-throughput experiments still remain costly and therefore experiments are usually carried out for a setting with far less samples than genes.  Shrinkage methods from high-dimensional statistics such as the graphical Lasso (Glasso) (\cite{Friedman01072008}) and the nodewise regression (\cite{meinshausen2006}) are a good choice for the graph reconstruction in this scenario. 
Gene regulatory networks contain regulatory genes with many interaction partners called \textit{hubs} that are essential for the viability of the organism because they are a central part of the interaction network (\cite{blais2005constructing}). This property also makes the hub genes potential drug targets.
Hub genes have been shown to be conserved in several organisms and it was suggested that their normal function is to act as genetic buffers, minimizing the effects of mutations in other genes (\cite{lehner2006systematic}). For example, the tumor suppressor Tp53 is a central hub in a molecular network that controls cell proliferation and death in response to oncogenic conditions (\cite{collavin2010p53}). The protein encoded by Tp53 is a powerful tumor suppressor, as proven by a studies of in vivo models and confirmed by frequent mutation in human cancers (\cite{donehower200920}). 
 It is very challenging to reconstruct such networks with hubs and the traditional $l_{1}$-based methods such as Glasso and nodewise regression perform poorly for these type of networks under the setting of less samples than genes. 
The edges of networks estimated from the standard $l_{1}$-norm based methods  correspond to a maximum posterior mode where independent, two sided exponential prior distribution are placed on the edge coefficients (\cite{hans2009bayesian}). Such approaches indirectly assume that each edge is treated equally which in turn corresponds to an Erd\"{o}s-R$\acute{\textnormal{e}}$nyi network in which most of the nodes have similar degrees (\cite{tan2014learning}, \cite{erdos1959random}).

A number of methods have been developed by different authors to tackle the problem of hub network inference from high-dimensional data. For example,  \cite{peng2009}  proposed a joint sparse regression model called SPACE (sparse partial correlation estimation) that allows to incorporate estimated degree information as a prior. However, their method does not perform well when the graph contains a few hubs that are highly connected to other nodes. 
\cite{liu2011learning} proposed a reweighted method to infer scale-free and hub type networks that performs better than graphical Lasso and nodewise regression. However, the networks considered in their study contain hub nodes that are far less connected than the hubs which we consider in our study. 
\cite{tan2014learning} proposed $l_{1}$-based method involving three penalty parameters that control hub sparsity, a selection of hub nodes and overall sparsity. The method performs well in the presence of highly connected nodes in the graph, however, it includes three penalty parameters and final estimates are highly dependent on how well these penalty parameters are chosen. 
In the context of Ising graphical models, \cite{tandon2014learning} provided theoretical guarantees for $l_{1}$-based logistic regression to recover the networks with a few hubs that have large degrees. Moreover, the authors provided a quantitative criterion to detect hub and non-hub nodes in the network. 
Other authors proposed methods to screen the hubs in the network in the context of graphical models (\cite{firouzi2013local}; \cite{hero2012hub}). However, these methods do not aim at estimating the hub network.

In this manuscript, we address the problem of estimating hub graphs from data in small $n$ large $p$ scenarios and propose a method (DW-Lasso) that consists of an iterative degree estimation step followed by a graph reconstruction step (Figure \ref{fig:01}). Our method in spirit is close to the method by \cite{tandon2014learning}, however, our method is designed for Gaussian graphical models and treats the genes as continuous variables.
We demonstrate the increased performance of our approach in comparison to traditional $l_{1}$-based methods under high-dimensional settings both on simulated and experimental data.

\section{Methods}
\newcommand{\indep}{\rotatebox[origin=c]{90}{$\models$}}
\subsection{Notation}
Consider a $p$-dimensional random vector $X = (X_{1}, \ldots, X_{p})^{T}$, which represents the expression levels of $p$ genes. We assume that $X$ follows the multivariate normal distribution with zero mean and covariance matrix $\boldsymbol{\Sigma}$.
We have a design matrix  $ \mathbf{X} = (\mathbf{X}_{1}, \ldots, \mathbf{X}_{p})$ of size $n \times p$, where 
$n$ is the number of observations that are  independently and identically distributed.\\
We consider a undirected graph $G = (V,E)$, with a set of nodes, $V = \{1, \ldots,p \}$ and a set of edges, $E \subseteq V \times V$. In our case, $V$ represents a set of genes and $E$ represents a set of conditional dependencies between the genes. \\
Under the multivariate normal distribution assumption, the concentration matrix $\boldsymbol{\Theta} \equiv \boldsymbol{\Sigma}^{-1}$ encodes the conditional independence relationships between genes through the graph $G$. In particular, the genes $i$ and $j$ are conditionally independent given the other genes, if $(i,j)$-th entry of the concentration matrix is zero. 

Let $\rho_{ij}$ denote a partial correlation  between any two genes $X_{i}$ and $X_{j}$. It is related to the elements of the concentration matrix through the relation
\begin{equation}\label{par_cor}
\rho_{ij} = -\dfrac{\Theta_{ij}}{\sqrt{\Theta_{ii}\Theta_{jj}}},
\end{equation}
for $i \neq j$ and $\rho_{ij} = 1$ for $i=j$. 
There is an edge between genes $i$ and $j$, if $\rho_{ij} \neq 0$  and no edge if $\rho_{ij} = 0$ for $i \neq j$.
Furthermore, we define the degree of gene $i$ by $d_{i}$ which represents the number of genes directly connected to gene $i$. It is computed using the following expression
\begin{equation}\label{degree}
d_{i} = \sum_{j=1, \\  j \neq i}^{p} \mathbb{1}{\{\Theta_{ij} \neq 0\}}, \forall i \in V.
\end{equation}

\begin{figure}[!t]%figure1
%	\centerline{\includegraphics[scale=0.5]{Figures/Figure1}}
		\centerline{\includegraphics[scale=0.5]{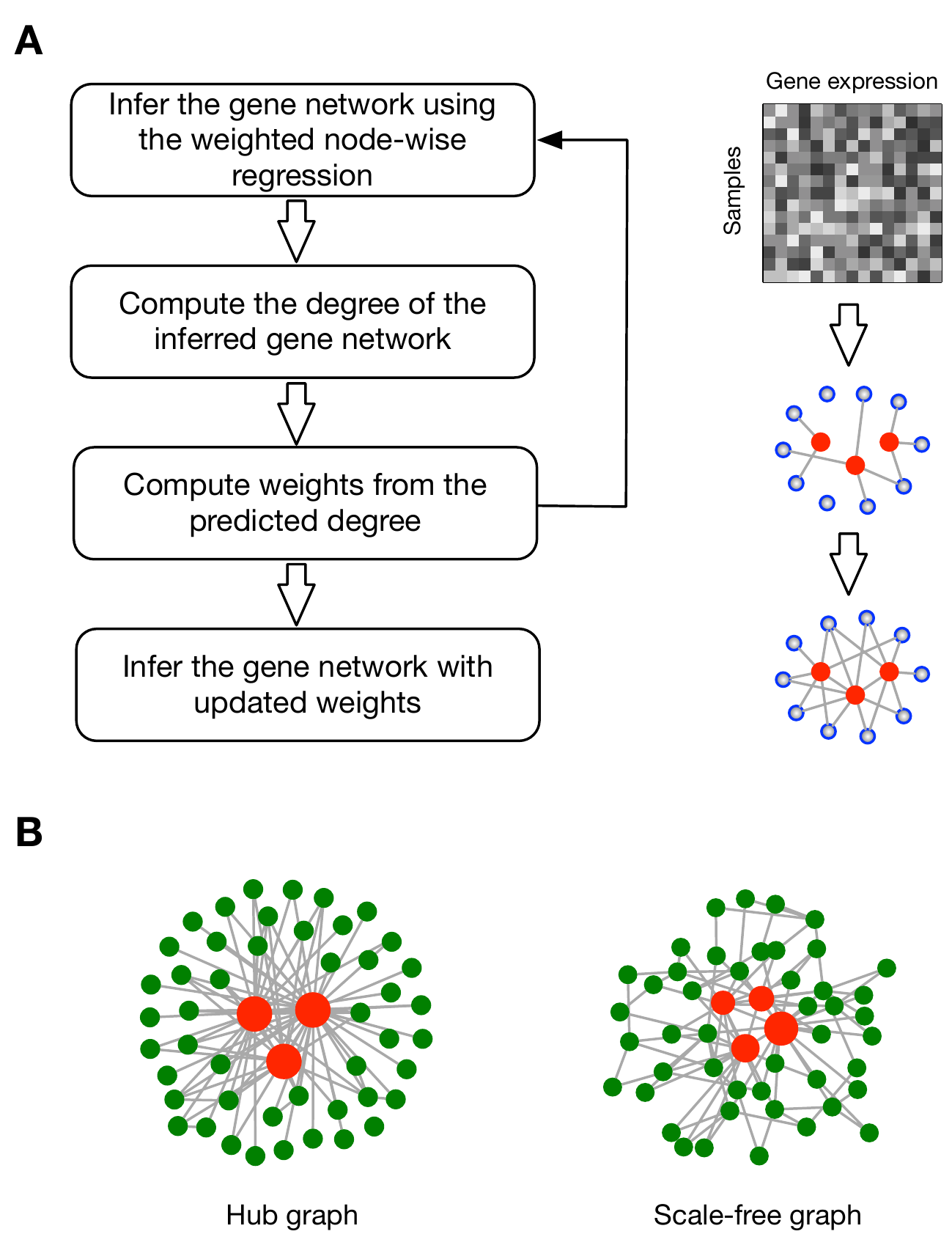}}
	\caption{{\small (A)  Workflow of the network inference with the degree weighted Lasso method (DW-Lasso).  Initially, the network is reconstructed using nodewise regression (MB-Lasso) which corresponds to the weighted nodewise-regression with the same weights. In the second step, the degree of the resulting network is computed. In the third step, the weights are computed from the estimated degrees. These weights are incorporated into  the weighted nodewise regression and updated until the algorithm attains a steady state. The weights at the steady state are then used to reconstruct the final network. Red nodes represent the hub nodes. (B) Illustration of hub and scale-free networks considered in the study.}}\label{fig:01}
\end{figure}

\subsection{Degree estimation using a nodewise regression method}
\cite{tandon2014learning} proposed a theoretical criterion that allows to differentiate between hubs and non-hubs.
This makes possible to infer the network using the non-hubs estimates by ignoring the estimates from hub nodes
under the assumption that the underlying network contains a few hubs with large degrees.
Motivating by this fact, we attempt to distinguish the hub genes based on their degrees using the nodewise regression approach (\cite{meinshausen2006}). For example, simulation results on scale-free networks indicate that nodewise regression is able to differentiate between hub and non-hubs nodes for some values of penalty parameter (Figure \ref{fig:13}A and for more information see Section \ref{sec:degree_separation}).  

For nodewise regression, we assume $\mathbf{X}_{i}$ is a response vector corresponding to gene $i$ and $\mathbf{X}^{\setminus i}$ is the submatrix after excluding the $i$-th column of the design matrix $\mathbf{X}$.
In order to reconstruct the graph, we assume a linear relationship between the genes. Therefore, we linearly regress gene $i$ given the remaining genes, using the following model 
\begin{equation}\label{regression}
\mathbf{X}_i = \mathbf{X}^{\setminus i}\boldsymbol{\beta}_{i} + \boldsymbol{\epsilon}_i,
\end{equation}
where $\boldsymbol{\beta}_{i}$ is a coefficient vector of length $p-1$ and 
$\epsilon_{i} \sim \mathcal{N}(0,\sigma^2\boldsymbol{I})$.
%$\boldsymbol{\epsilon}_{i}$ is the noise term in the model. 
%with$\mathbb{E}[\boldsymbol{\epsilon}_{i}]=\mathbf{0}$.  
Each element of the vector $\boldsymbol{\beta}_i$ can be expressed in terms of partial correlations by 
\begin{equation}\label{par_cor2}
\beta_{ij}=   -\rho_{ij} \sqrt{\dfrac{\Theta_{jj}}{\Theta_{ii}}},
\end{equation}
where $\rho_{ij}$ is defined in \eqref{par_cor}. Under the high-dimensional setting ($p>n$), the coefficient vector $\boldsymbol{\beta}_i$ cannot be uniquely estimated, because $(\mathbf{X}^{\setminus i})^{T}\mathbf{X}^{\setminus i}$ is not invertible.
Therefore, we adopt a Lasso approach (\cite{tibshirani1996regression}) to estimate the coefficient vector for gene $i$ using the following objective function
\begin{equation}\label{lasso}
\hat{\boldsymbol{\beta}}_{i} = \arg\min_{\boldsymbol{\beta}_{i}}\left(\frac{1}{n}||\mathbf{X}_{i} - \mathbf{X}^{\setminus i} \boldsymbol{\beta}_{i}||^{2}_{2} + \lambda_{1}||\boldsymbol{\beta}_{i}||_{1}\right),
\end{equation}
where $\lambda_{1} >0$ is the penalty parameter that enforces sparsity in the estimates. 
By performing a nodewise regression for all genes (\cite{meinshausen2006}), we estimate the off-diagonal elements of the concentration matrix $\hat{\Theta}_{ij}=\hat{\beta}_{ij}$ for all $i \neq j$. Since the estimates of \eqref{lasso} are not symmetric, we apply the following transformation to obtain the symmetric matrix
\begin{equation}\label{eq:transformation}
\tilde{\boldsymbol{\Theta}} = (\hat{\boldsymbol{\Theta}}+\hat{\boldsymbol{\Theta}}^T)/2,
\end{equation}
Furthermore,  we compute the  degree vector $\boldsymbol{\hat{d}}=(\hat{d}_{1},\ldots,\hat{d}_{p})$ from the estimated concentration matrix  $\tilde{\boldsymbol{\Theta}}$, where each element of the vector $\boldsymbol{\hat{d}}$ is defined as
\begin{equation}\label{degree}
\hat{d}_{i} = \sum_{j=1, \\  j \neq i}^{p} \mathbb{1}{\{\tilde{\Theta}_{ij} \neq 0\}}, \forall i \in V
\end{equation}
In the next step, we compute a weight for gene $i$ by
\begin{equation}\label{eq:weight1}
w_{i}=(1+\hat{d}_{i})^{-1}.
\end{equation}
When the estimated degree of a gene $i$ is zero, then the corresponding weight is equal to one. 
A similar formulation has been applied for the adaptive Lasso using the weight as a function of estimated coefficients (\cite{zou2006adaptive}). 
Furthermore, we normalize the weights to represent them as probabilities 
\begin{equation}\label{eq:norm_weight}
\bar{w}_{i}=\dfrac{w_{i}}{\sum_{j}w_{j}}, \ \forall i \in V
\end{equation}

Finally, we exploit the weight vector iteratively $\boldsymbol{\bar{w}}=(\bar{w}_{1},\ldots,\bar{w}_{p})$
 as prior information to reconstruct the gene-gene interaction graph using the weighted Lasso.

\subsection{Iterative degree estimation and network reconstruction using a weighted nodewise regression method}\label{iterative_section}
\begin{figure}[!tpb]%figure1
%	\centerline{\includegraphics[scale=0.4,bb=0 0 532 237]{Figures/Figure2}}
		\centerline{\includegraphics[scale=0.4,bb=0 0 532 237]{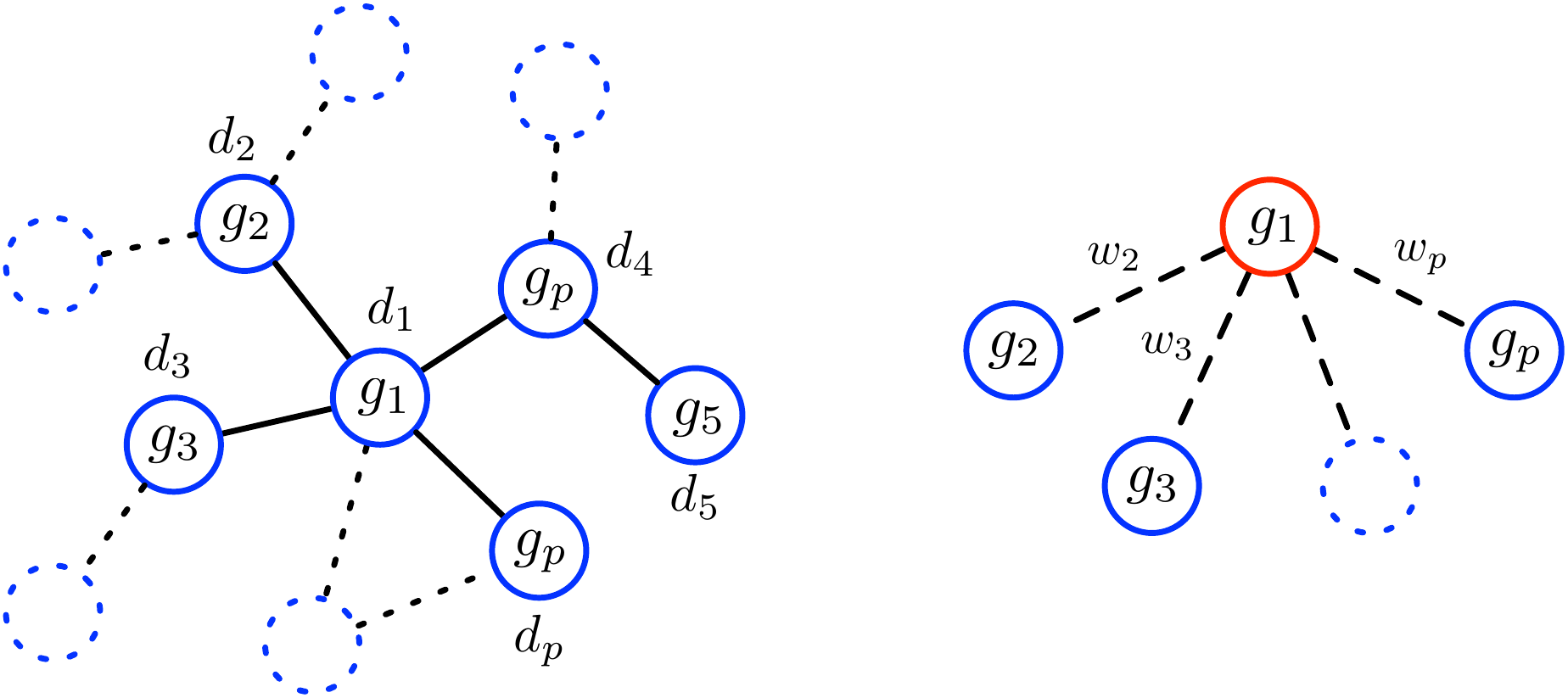}}
	\caption{{ \small Integrating degree information of the graph into a Lasso setting. Illustration of the network with degree information $d_{i}, \ i=1,\ldots,p$ (left).  We show how the degree information  is integrated into a weighted Lasso setting for gene 1 (right). The corresponding weighted Lasso for gene 1 is then defined as $\hat{\boldsymbol{\beta}}_{1} = \arg\min_{\boldsymbol{\beta}_{1}}\left(n^{-1}||\mathbf{X}_{1} - \mathbf{X}^{\setminus 1} \boldsymbol{\beta}_{1}||^{2}_{2} + P(\boldsymbol{\beta}_{1}) \right)$ with $P(\boldsymbol{\beta}_{1})=|w_{2}\beta_{12}| + |w_{3}\beta_{13}| + \cdots + |w_{p}\beta_{1p}|)$.}}\label{fig:02}
\end{figure}

We propose an iterative method to update the degree vector using the current solution. A similar iterative approach was proposed earlier by \cite{candes2008enhancing}, where the weight vector is computed from the current value of regression coefficients.
By employing a weighted Lasso setup, we formulate the iterative approach as follows: 
\begin{equation}\label{lasso_iter}
\hat{\boldsymbol{\beta}}_{i}^{(k+1)} = \arg \min_{\boldsymbol{\beta}_{i}^{(k+1)}}\dfrac{1}{n}||\mathbf{X}_{i}-\mathbf{X}^{\setminus i}\boldsymbol{\beta}_{i}^{(k+1)}||_{2}^{2} + P_{k}(\boldsymbol{\beta}_{i}^{(k+1)})
\end{equation}
where 
\begin{equation}
\begin{split}
P_{k}(\boldsymbol{\beta}_{i}^{(k+1)}) = %\lambda_{1}||\bar{\boldsymbol{w}}_{-i}^{(k)}\boldsymbol{\beta}_{i}^{(k+1)}||_{1} \\
\lambda_{1}\sum_{\ j \neq i}\bar{w}_{j}^{(k)}|\beta_{ij}^{(k+1)}|  
\end{split}
\end{equation}
where, $\bar{w}_{j}^{(k)}$ is  the normalized weight for gene $j$ at $k$-th iteration. For $k=1$, we initialize the weight vector as a normalized unit vector and  the problem \eqref{lasso_iter} reduces to the standard nodewise regression method defined in \eqref{lasso}.
For any gene $j$, if the weight $\bar{w}_{j}^{(k)}$ is small, the coefficient $\beta_{ij}^{(k+1)}$ is less penalized and this leads to a recovery of the edge between genes $i$  and $j$. 
Suppose, in the true graph, there is an edge between a  high degree gene $i$ and a low degree gene $j$. The coefficient $\beta_{ij}^{(k+1)}$ shrinks to zero with high probability, due to high weight $\bar{w}_{j}^{(k)}$. In contrast, the coefficient $\beta_{ji}^{(k+1)}$ is recovered with high probability due to low weight $\bar{w}_{i}^{(k)}$. This introduces a further asymmetry in the graph reconstruction. We obtain symmetric estimates using the transformation \eqref{eq:transformation}.
The illustration of how each gene is weighted by the degree of other genes is depicted in a sample graph with four nodes in Figure \ref{fig:02}. 
We update the normalized weight vector according to the following recursion
\begin{equation}
\begin{split}
\bar{\boldsymbol{w}}^{(k)}  = \alpha \bar{\boldsymbol{w}}^{(k-1)} + (1-\alpha) \boldsymbol{\Psi}(\bar{\boldsymbol{w}}^{(k-1)}), \ k \geq 1, \\
\end{split}
\end{equation}
where the $i$-th component of vector $\boldsymbol{\Psi}$ reads 
\begin{equation}
\begin{split}
\Psi_i(\bar{\boldsymbol{w}}^{(k-1)})  = \dfrac{(1+\hat{d}^{(k-1)}_{i})^{-1}}{\sum_{j}1/(1+\hat{d}^{(k-1)}_{j})},
\end{split}
\end{equation}
with $\alpha$ a constant such that $0<\alpha<1$. 
We compute the degree vector $\boldsymbol{\hat{d}}^{*}$ via \eqref{degree} at the $(k+1)$-th iteration using the estimated coefficients from \eqref{lasso_iter}  and using the transformation \eqref{eq:transformation}.
Using \eqref{eq:weight1} and \eqref{eq:norm_weight},  we then obtain the normalized weight vector  $\boldsymbol{\bar{w}}^{*}=(\bar{w}^{*}_{1},\ldots,\bar{w}^{*}_{p})$,
which is used  to reconstruct the final graph employing a weighted Lasso regression
\begin{equation}\label{final_lasso}
\hat{\boldsymbol{\beta}}_{i} = \arg\min_{\boldsymbol{\beta}_{i}}\left(\frac{1}{n}||\mathbf{X}_{i} - \mathbf{X}^{\setminus i} \boldsymbol{\beta}_{i}||^{2}_{2} + P(\boldsymbol{\beta}_{i}) \right),
\end{equation}
with
\begin{equation}
P(\boldsymbol{\beta}_{i}) = 
\lambda_{2}\sum_{j \neq i}\bar{w}_{j}^*|\beta_{ij}|,
\end{equation}
where $\lambda_{2} >0$ is the penalty parameter that controls the sparsity in the estimates.
In Section \ref{sec:convergence}, for a carefully chosen $\lambda_{1}$, we demonstrate that weight estimates converge close to weights defined by the true network. Additionally, we show that the performance of DW-Lasso increases with the number of iterations and achieves best results in terms of AUROC at the steady state (Figure \ref{fig:13}C).

\subsection{Justification of degree separation for nodewise regression}\label{sec:degree_separation}
Lets assume that there are $r$ hub and $p-r$ non-hub genes in the network.
%Lets denote the hub genes by $Z^{h}_{j}$, $j=1,\ldots,m$ and non-hub genes by $Z^{nh}_{r}$, $r=m+1,\ldots, p$. 
Lets denote the degree of hub and non-hub genes  by $\bar{d}_{i}$, $i=1,\ldots,r$ and $\tilde{d}_{j}$, $j=r+1,\ldots, p$,  respectively.
Assume that $\bar{d}_{i}>n, \ i=1,\ldots,r$ and $\tilde{d}_{j} <n, \ j=r+1,\ldots, p$ such that for hub genes the Lasso estimation involves more coefficients than samples to estimate, while for non-hub genes it involves less coefficients than the samples.  In case of hub genes, it leads to an underestimation problem, meaning that some coefficients cannot be uniquely estimated. 
In contrast, for non-hubs, the coefficients can be uniquely estimated using large penalty parameters. If we assume that each hub node is connected to most non-hub nodes, some edges connected to hubs can be recovered by the edges estimated from non-hub nodes.
Therefore, the nodewise regression is able to partially infer the edges connected to hub nodes where the number of false positive edges is controlled for large penalty parameters.
Simulations with nodewise regression demonstrate that the degree of hub and non-hubs can be separated for the intermediate values of the penalty parameter (Figure \ref{fig:13}A).

\subsection{Extending the methodology to transcription factor-gene interaction networks}\label{sec:DW_Lasso_TF}
The proposed methodology can be extended to reconstruct transcription factor-gene interaction networks. In this case, we are only interested in the inference of edges between transcription factors and genes. 
We consider a set of transcription factors $Q =\{1,\ldots,m\}$, $\ Q \subset V$ and a set of genes $G =\{m+1,\ldots,p\}$, $\ G \subset V$. Let $\mathbf{X}^Q$ be a $n \times m$-matrix that represents the expression level of $m$ transcription factors. Similarly, let $\mathbf{X}^G$ be a $n \times (p-m)$-matrix that represents the expression level of $p-m$ genes.
To reconstruct the network of transcription factors and genes, we linearly regress  gene $i$ against all transcription factors
\begin{equation}
\mathbf{X}^{G}_{i} = \mathbf{X}^{Q}\boldsymbol{\beta}^Q_{i} + \boldsymbol{\epsilon}_{i}^G
\end{equation}
where $\boldsymbol{\beta}^Q_{i}$ is a coefficient vector of length $m$ and $\boldsymbol{\epsilon}_{i}^G \sim \mathcal{N}(0,\sigma^2\boldsymbol{I})$.
We estimate $\boldsymbol{\beta}^Q_{i}$ using the Lasso
\begin{equation}\label{lasso_TF}
\hat{\boldsymbol{\beta}}_{i}^{Q} = \arg \min_{\boldsymbol{\beta}_{i}^{Q}}\dfrac{1}{n}||\mathbf{X}_{i}^G-\mathbf{X}^{Q}\boldsymbol{\beta}_{i}^{Q}||_{2}^{2} + ||\boldsymbol{\beta}_{i}^{Q}||_{1}
\end{equation}
By regressing all genes against transcription factors, we estimate the $\hat{\boldsymbol{\Theta}}^{QG}$ - $(p-m) \times m$ matrix that includes all pairwise estimates between genes and transcription factors. We can further compute the degree of transcription factor $j$ using \eqref{degree}
\begin{equation}\label{degree_TF}
\hat{d}_{j}^{Q} = \sum_{i=1}^{p-m} \mathbb{1}{\{\hat{\Theta}_{ij}^{QG} \neq 0\}}, \forall j
\end{equation}
By using \eqref{eq:weight1}, the degree $\hat{d}_{j}^{Q}$ can be integrated into a weighted Lasso setting
\begin{equation}\label{lasso_weight_TF}
\hat{\boldsymbol{\beta}}_{i}^{Q} = \arg \min_{\boldsymbol{\beta}_{i}^{Q}}\dfrac{1}{n}||\mathbf{X}_{i}^G-\mathbf{X}^{Q}\boldsymbol{\beta}_{i}^{Q}||_{2}^{2} + P(\boldsymbol{\beta}_{i}^{Q})
\end{equation}
with
\begin{equation}
\begin{split}
P(\boldsymbol{\beta}_{i}^{Q}) = %\lambda_{1}||\bar{\boldsymbol{w}}_{-i}^{(k)}\boldsymbol{\beta}_{i}^{(k+1)}||_{1} \\
\lambda_{1}\sum_{j=1}^{m}\bar{w}_{j}^{Q}|\beta_{ij}^{Q}|  
\end{split}
\end{equation}
%where, the normalized weight vector of $(k-1)$th iteration is given as
where, $\bar{w}_{j}^{Q}$ is  the normalized weight for transcription factor $j$.
These steps can be solved iteratively as described in Section \ref{iterative_section}. 

\subsection{Choosing penalty parameters}\label{sec:penalty_selection}
In this section, we discuss how to select the penalty parameter $\lambda_{1}$ for the degree estimation step. 
%Tuning these both parameters allow to explore a broad solution space, hence it is very challenging to choose the values close to optimal ones. 
Therefore, we propose to employ a stability selection (\cite{meinshausen2010stability}) to choose the penalty parameters from the data .  
According to \cite{meinshausen2010stability}, we denote a set of estimated graphs as $\hat{G}^{\lambda}=\{(i,j)\in E; \hat{\boldsymbol{\Theta}}_{ij}\neq0 \}$ obtained for every value $\lambda \in \mathbb{R}_{+}$. Denote a set of samples as $I_{n}=\{1,\ldots,n\}$. Let $I$ be a random subsample of $I_{n}$ of size $n/2$ drawn without replacement. 
If a random subsample $I$ is drawn $J$ times, for every set $D \subseteq E$,  one can define the probability of being in the selected set $\hat{G}^{\lambda}$
\begin{equation}\label{prob_frequency2}
\hat{\Pi}_{D}^\lambda = J^{-1}\sum_{j=1}^{J}\mathbb{1}\{D\subseteq\hat{G}^{\lambda}(I^j) \}
\end{equation}

For a set of regularization parameters $\Lambda$ and a cutoff $\pi_{thr}$, the set of stable edges is defined as 
\begin{equation}
\hat{G}^{stable}=\{D: \max_{\lambda \in \Lambda}\hat{\Pi}_{D}^\lambda \geq \pi_{thr} \}, \ \pi_{thr} \in (0,1)
\end{equation}
where the edges with high selection probabilities are selected, while the edges with low selection probabilities are excluded. As justified in  \cite{meinshausen2010stability},  for $\pi_{thr} \in (0.5,1]$, it is possible to control the number of falsely selected edges $W$ given the expected number of selected edges $q_{\Lambda}$, the total number of edges $\bar{q}$ and the threshold parameter $\pi_{thr}$
\begin{equation}\label{error_control}
\mathbb{E}(W) \leq \dfrac{q_{\Lambda}^2}{\bar{q}(2\pi_{thr}-1)}
\end{equation}
under the assumption that the DW-Lasso performs better than the random guess (exchangeability condition, for more information see \cite{meinshausen2010stability}). 
For a user defined $\pi_{thr}$ and given number of maximum false positives $\mathbb{E}(W)$, we compute the expected number of edges $q_{\Lambda}$ in the graph. Using the computed value of $q_{\Lambda}$, we select the penalty parameter $\lambda_{1}$.

 \subsection{Graph generation procedure}\label{sec:graph_generation}
 To generate a hub graph, we  generate a sparse symmetric adjacency matrix  $\boldsymbol{A} \in \mathbb{R}^{p \times p}$  with the edge probability  $p_{1}$. We next randomly select a set of hub nodes of size $h$  which is a predefined parameter. Finally, we generate a set of neighboring nodes around hubs with probability $p_{2}$. The parameters $p_{1}$ and $p_{2}$ allow us to generate various hub graphs with different sparsity levels. In our case, we set $p_{1}=0.01$ and $p_{2}=0.95$.
 
 We also evaluate the performance of our method on scale-free graphs (\cite{Barabasi15101999}, \cite{durrett2007random}).
 We use a linear preferential attachment approach to generate an adjacency matrix  $\boldsymbol{A} \in \mathbb{R}^{p \times p}$ for a scale-free graph with degree exponent $\gamma$. 
 The approach starts with a connected graph that contains $m_{0}$ nodes. The new nodes with $m \leq m_{0}$ edges are added to $m_{0}$ existing nodes in the graph. A new node is added to the existing node $i$ depending on the degree $k_{i}$ with  the probability $P(k_{i}) = k_{i}/\sum_{j}^{}k_{j}$. In our case, we set $m=3$.

 Given the adjacency matrix $\boldsymbol{A} \in \mathbb{R}^{p \times p}$, we uniformly generate a new weighted adjacency matrix  
 $$
 B_{ij} \stackrel{\text{i.i.d}}{\sim} \begin{cases} \textnormal{Unif}([-0.75,-0.25]\cup[0.25,0.75]), & \mbox{if } A_{ij} \neq 0 \\ 0, & \mbox{if } A_{ij}=0 \end{cases}
 $$
 Finally, we convert $\boldsymbol{B}$ into a symmetric matrix \\
 $\bar{\boldsymbol{B}} = (\boldsymbol{B}+\boldsymbol{B}^T)/2$.
 We transform the matrix $\bar{\boldsymbol{B}}$ to the concentration matrix using the expression 
 \begin{equation}
 \boldsymbol{\Theta}=\bar{\boldsymbol{B}} + (0.1 + |\lambda_{\min}(\bar{\boldsymbol{B}})|)\boldsymbol{I},
 \end{equation}
 where $\lambda_{\min}(.)$ is the smallest eigenvalue of the matrix.
 %Eventually, the weighted adjacency matrix $\bar{B}$ is converted to the concentration matrix $\Theta \succ 0$ which is positive definite.
 
 Given the concentration matrix $\boldsymbol{\Theta}$, 
 we generate samples that follow a multivariate Gaussian distribution with mean $\boldsymbol{0}$ and covariance matrix $\boldsymbol{\Sigma}=\boldsymbol{\Theta}^{-1}$.
 In the final step, the data is standardized to mean 0 and standard deviation $\sigma=1$.

\subsection{Quantifying the functional content of a graph}\label{sec:genesets_metrics}
To quantify the association between publically available gene sets and the inferred networks, we compute two metrics: Knet score implemented in SANTA package (\cite{cornish2014santa}) that takes into account the global topology of the network, and compactness score that quantifies the average distance between the genes in the network (\cite{glaab2010extending}). 
Knet function is a modified form of Ripley's K-function (\cite{gaetan2010spatial}) and is defined as
\begin{equation}
K^{net}(s)=\dfrac{2}{(\bar{k}n)^2}\sum_{i}k_{i}\sum_{j}(k_{j}-\bar{k})I (d(i,j) \leq s)
\end{equation} 
where, $k_{i}$ is the phenotype observed at gene $i$,  $\bar{k}=\dfrac{1}{n}\sum_{i=1}^{n}k_{i}$, $d(i,j)$ is the shortest distance between two hits in the network,  and $I (d(i,j) \leq s)$ is the indicator function which equals 1, if the distance $d(i,j)$ between hits $i$ and $j$ is smaller or equal to $s$, and 0 otherwise.
For a given gene set $M$,  the compactness score is defined as
\begin{equation}
CS(M) =  \dfrac{2\sum_{i,j \in M; i <j}d(M_{i},M_{j})}{|M|(|M|-1)}
\end{equation}
where $d(M_{i},M_{j})$ is the shortest distance between any two pairs of genes in gene set $M$, and $|M|$ is the cardinality of gene set $M$.

\begin{figure*}[!t]%figure1
%	\centerline{\includegraphics[scale=0.28]{Figures/Figure3}}
		\centerline{\includegraphics[scale=0.28]{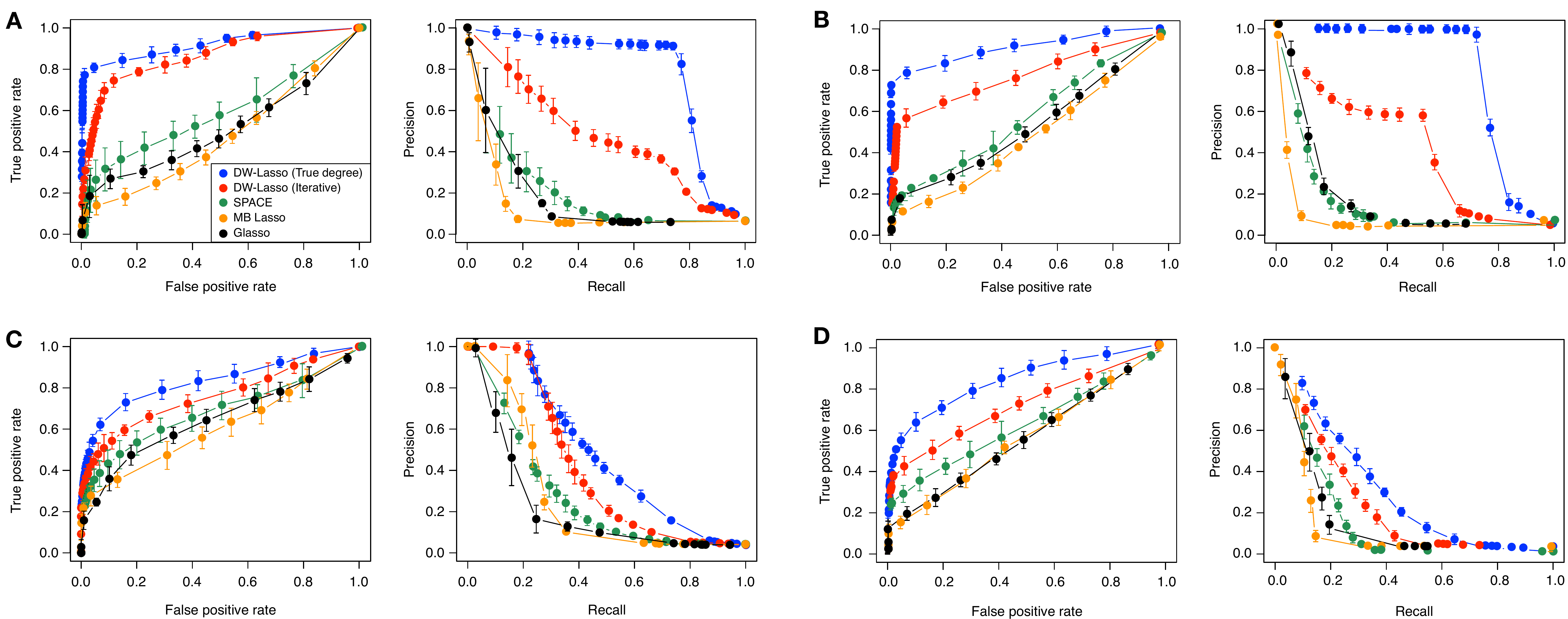}}
	\caption{{ \small Performance of various methods on (A)-(B) hub and (C)-(D) scale-free networks in the simulation study ((A)-(C) $p=100, \ n=50$ and (B)-(D) $p=500, \ n=100$). The considered methods are DW-Lasso with true degree information and the proposed iterative procedure, sparse partial correlation estimation (SPACE) (\cite{peng2009}), graphical Lasso (Glasso) (\cite{Friedman01072008}) and nodewise regression (MB-Lasso) (\cite{meinshausen2006}). ROC and PR curves for DW-Lasso are computed for fixed values of $\lambda_{1}$ and varying $\lambda_{2}$. For hub networks, the penalty is chosen as $\lambda_{1}=0.65$  and  for scale-free networks as $\lambda_{1}=0.45$ (selected by stability selection). Degree estimates at $k=30$ iterations are used for the network inference. ROC and PR curves for Glasso and nodewise regression are computed by varying the penalty parameters.  Error bars represent two standard deviation of the mean.}}\label{fig:03}
\end{figure*}

\section{Results on synthetic data}
\subsection{Performance assessment}
We consider the following metrics to evaluate the performance of our method in comparison to state-of-the- art methods.

\begin{itemize}
	\item Receiver operation characteristics (ROC): \\
	$\textnormal{TPR}=\dfrac{\textnormal{TP}}{\textnormal{TP}+\textnormal{FN}}$, $\textnormal{FPR}=\dfrac{\textnormal{FP}}{\textnormal{FP}+\textnormal{TN}}$
	\item Precision and Recall (PR):\\
	$\textnormal{Precision}=\dfrac{\textnormal{TP}}{\textnormal{TP}+\textnormal{FP}}$, $\textnormal{Recall}=\dfrac{\textnormal{TP}}{\textnormal{TP}+\textnormal{FN}}$ 
	\item Area under the ROC curve (AUROC)
	\item Area under the Precision and Recall curve (AUPR)
\end{itemize}	
ROC and PR curves are computed for a fixed value of $\lambda_{1}$ which has been selected by stability selection (\cite{meinshausen2010stability}) (Section \ref{sec:penalty_selection}) and  by varying the penalty parameter $\lambda_{2}$ and AUROC and AUPR  have been computed from corresponding ROC and AUPR curves.  

\subsection{Performance on synthetic data} 
We first evaluate the performance of our method in comparison to state-of-the-art methods such as nodewise regression (MB-Lasso) (\cite{meinshausen2006}), graphical Lasso (Glasso); (\cite{Friedman01072008}) and sparse partial correlation estimation (SPACE) (\cite{peng2009}), based on \textit{in silico} study. 
%We optimize the predicted graphs based on the true ones, and then compare the performance of all methods together.
The performance of selected methods is assessed on hub and scale-free graphs (Figure \ref{fig:01}B). 
We generate a set of hub networks using the procedure described by \cite{tan2014learning}. We generate scale-free networks using the procedure given in (\cite{Barabasi15101999}; \cite{durrett2007random}).
Detailed information can be found in Section \ref{sec:graph_generation}.

\begin{figure*}[t]%figure1
	%\centerline{\includegraphics[width=\textwidth]{Figures/roc_pr_hub_p100_n50_hn3_sp4}}
%	\centerline{\includegraphics[scale=0.25]{Figures/Figure4}}
		\centerline{\includegraphics[scale=0.25]{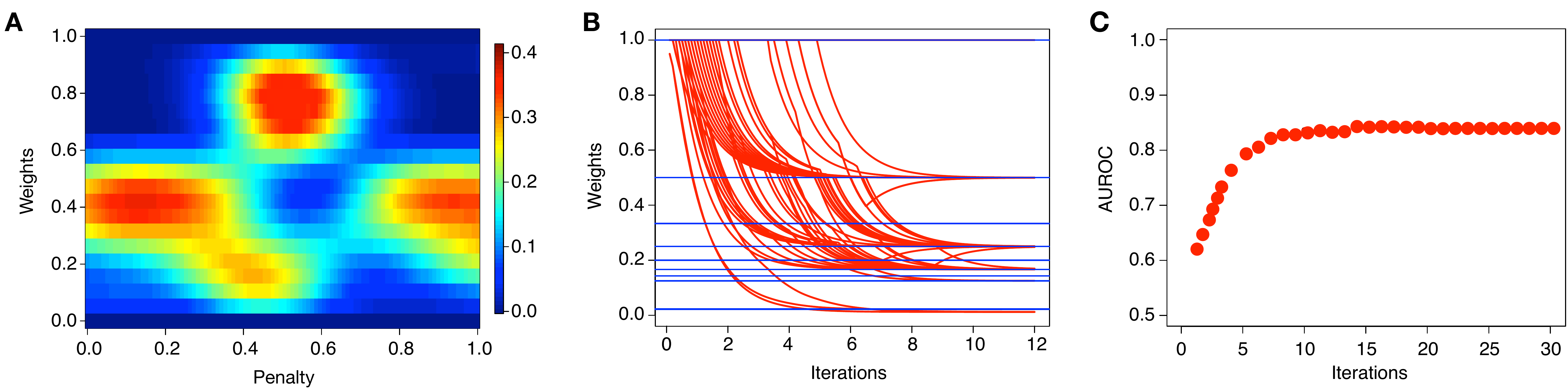}}
	\caption{{ \small (A) Heatmap of weights computed for different values of penalty parameter using nodewise regression. Initially for small penalty values, all genes have similar weights. For intermediate penalty values, there is a separation of weights between hub and non-hub genes (indicates separation of degrees between hubs and non-hubs). For large penalty values, the network is very sparse and hence all genes have the similar weights. (B), (C) The convergence of the weight $w^{(k)}$ for DW-Lasso method demonstrated for $p=50$, $n=40$, $\lambda_{1}=0.5$. (B) The trajectories of weights (red lines) over different iterations that converge to the steady state. Blue lines indicate the true weights computed from the true network. Initially, all weights are same and equal to one. (C) The prediction accuracy of the DW-Lasso increases with the number of  iterations.}}\label{fig:13}
\end{figure*}

We randomly create hub networks with node sizes $p=100$ and $500$ that contain three and ten hub nodes, respectively. Here, we consider an extreme setting where the hub nodes are connected to 95 $\%$ of non-hub nodes. Moreover, scale-free networks are also generated with node sizes of $p=100$ and $500$. We generate scale-free networks with degree exponent $\gamma=2.5$, which also contain hub nodes.
Furthermore, we generate 20 different datasets of sample sizes $n=50$ for $p=100$ and $n=100$ for $p=500$ both for hub and scale-free networks.
This allows us to evaluate the performance of the methods in presence of sample noise.
 The DW-Lasso achieves a good performance in the large $p$ and small $n$ setting (Figure \ref{fig:03}),  when the true degree information is supplied as a weight vector, which highlights the importance of degree information on the predictive power.  

 Therefore, the predictions using the true degree information give the upper bound for the prediction accuracy of the proposed method.
 Simulation results demonstrate that our DW-Lasso performs better than the  state-of-the-art methods. In the case of hub networks, all three methods perform poorly, whereas DW-Lasso detects most of the true positives successfully, even when the number of samples is less than the number of genes (Figure \ref{fig:03}A and Figure \ref{fig:03}B). In the setting $p=500$ and $n=100$, the performances of the state-of-the-art methods are almost comparable to a random guess, while DW-Lasso performs far better than the random guess with AUROC greater than $0.7$ (Figure \ref{fig:03}B). 
In the case of scale-free networks, the DW-Lasso also performs better than the competing methods (Figure \ref{fig:03}C and Figure \ref{fig:03}D). 
%Corresponding AUROC and AUPR scores with different samples are given in Table 1. 
Further simulation results show that DW-Lasso does not perform well for traditional scale-free networks with degree exponent $\gamma=3$, which results in networks that are extremely sparse. In these networks,  the number of edges is roughly equal to the number of genes and the graph can be efficiently reconstructed using the traditional methods.

\begin{figure}[tb]%figure1
	%\centerline{\includegraphics[width=\textwidth]{Figures/roc_pr_hub_p100_n50_hn3_sp4}}
%	\centerline{\includegraphics[scale=0.27]{Figures/Figure5}}
		\centerline{\includegraphics[scale=0.27]{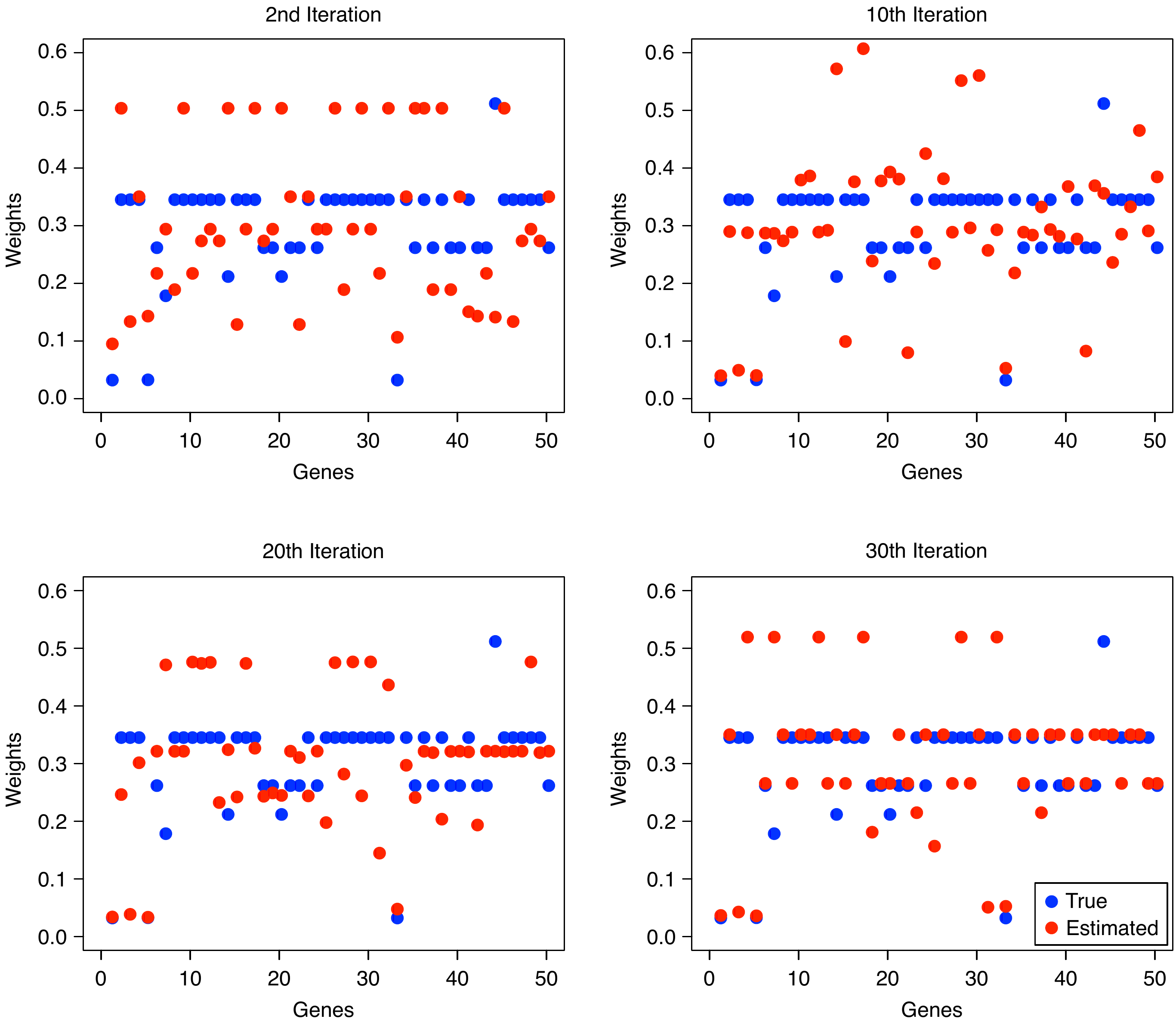}}
	\caption{{ \small The convergence of the degree weighted Lasso (DW-Lasso). Blue circles indicate the true weights; red circles indicate the predicted weights. The plots indicate that weights converge closely to true weigths at later iterations.}}\label{fig:08}
\end{figure}

\subsection{Convergence of degree estimates of DW-Lasso}\label{sec:convergence}
One main feature of DW-Lasso is the convergence of degree (weight) estimates. For a carefully chosen $\lambda_{1}$ as described in Section \ref{sec:penalty_selection}, most of the time, the DW-Lasso weight estimates converge close to the weights determined by the true graph (Figure \ref{fig:13}B and Figure \ref{fig:08}). The performance of DW-Lasso improves with the number of iterations and achieves the best estimates in terms of AUROC at steady state (Figure \ref{fig:13}C). For prediction accuracy, the convergence of the estimates is highly dependent on the choice of the penalty parameter $\lambda_{1}$. In particular, if $\lambda_{1}$ is chosen outside of the high performance region (see Section \ref{sec:influence}), the weight estimates of DW-Lasso converge to different steady states with low prediction accuracy.

\subsection{Effect of the number of hub genes in the network}
We next investigate how the number of hub genes in the network affects the performance of DW-Lasso. Therefore, we consider the network with $p=100$ genes  and create 20 hub networks with different numbers of hubs ($95 \%$ connectivity) that constitute $2 \%$ to $20 \%$ of the total genes. We randomly generate datasets of size $n=50$ and evaluate the performance of the aforementioned methods using  AUROC and AUPR scores (Figure \ref{fig:09}A). Simulation results demonstrate that the DW-Lasso outperforms other methods both in AUROC and AUPR scores throughout the considered hub densities. SPACE performs better than Glasso and nodewise regression in AUROC, but exhibits a similar performance in AUPR. As the number of hub genes increases, the performance of DW-Lasso decreases based on AUROC scores, which is also the case for SPACE. Glasso and nodewise regression perform as poorly as the random guess for large number of hub genes. The increase in AUPR score of the state-of-the-art methods, as the number of hub genes increases is trivial, since the sparsity of the graph decreases, which leads to less false positives in the estimates. While the AUPR scores of these methods increase, they perform as poorly as the random guess.

\subsection{Influence of penalty parameters}\label{sec:influence}
\begin{figure}[t]%figure1
	%\centerline{\includegraphics[width=\textwidth]{Figures/roc_pr_hub_p100_n50_hn3_sp4}}
%	\centerline{\includegraphics[scale=0.25]{Figures/Figure6}}
		\centerline{\includegraphics[scale=0.25]{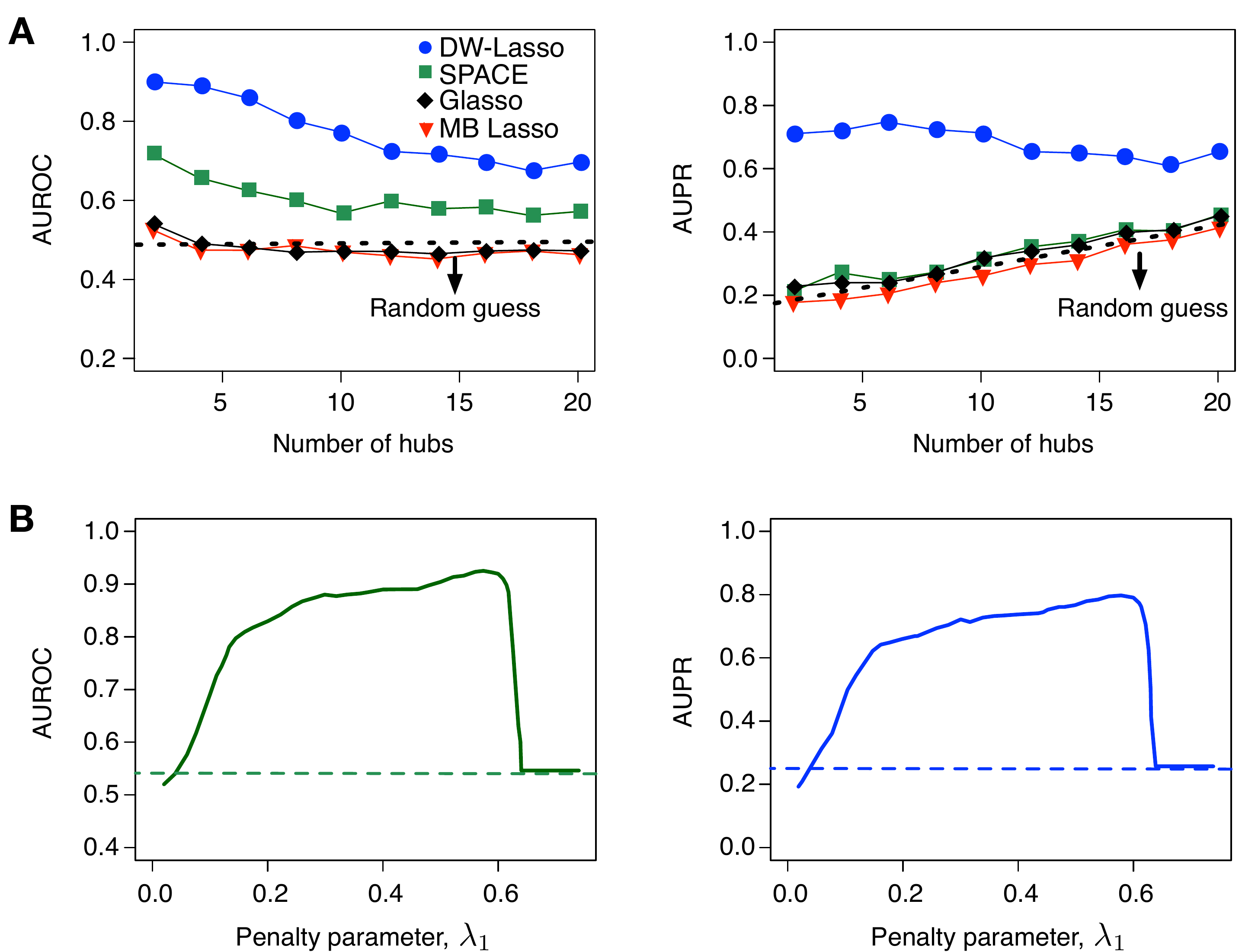}}
	\caption{{\small Effect of the different number of hubs and penalty parameters on the performance of DW-Lasso and  competing methods. (A) Effect of different hubs on the performance of selected methods. The simulations are performed for hub networks with different number of hubs in the setting $p=100, \ n=50$. Black dashed lines indicate random guess.
		(B) The influence of the penalty parameter $\lambda_{1}$ on the inferred networks.  Selection of the penalty parameter $\lambda_{1}$ is based on the AUROC and AUPR computed by varying $\lambda_{2}$ (computed for $k=30$ iterations). The optimal AUROC and AUPR is attained for the intermediate values of the penalty parameter. Dashed green and blue lines indicate the AUROC and AUPR scores obtained with the nodewise regression.}}\label{fig:09}
\end{figure}

Given the two penalty parameters in our method, we conduct simulations with hub graphs under the setting $p=60$ and $n=30$, where for each fixed value of  $\lambda_{1}$, we vary $\lambda_{2}$ and compute AUROC and AUPR scores. Simulation results demonstrate that the optimal AUROC and AUPR scores compared to those of the nodewise regression (dashed line) are obtained for intermediate values of $\lambda_{1}$ (Figure \ref{fig:09}B). We call this region the high performance region. The penalty parameter $\lambda_{1}$ chosen via stability selection falls in this region. In this high performance region, the number of false positive edges in the inferred graph is controlled such that the degree information estimation includes less false positive edges.
When $\lambda_{1}$ is large enough, such that the resulting graph is very sparse, the predictions of DW-Lasso is similar to those of the nodewise regression method.
Another distinct feature of our method is that it is able to infer graphs with different hub degree controlled by $\lambda_{1}$. When one increases $\lambda_{1}$, most of the edges tend to accumulate to a few hub genes, which are also known as super hubs (\cite{hao2012revisiting}). In contrast, decreasing $\lambda_{1}$ leads to graphs containing more hubs of lesser degree.

\section{Results on real data}
\subsection{Kidney Cancer gene expression data}
We apply our method to gene expression datasets from TCGA datasets for Kidney Clear Cell Carcinoma (KIRC) measurements (\cite{mclendon2008comprehensive}). We downloaded the normalized RNA sequencing dataset from UCSC Xena, a data server-based platform that stores functional genomics data (\cite{goldman2016ucsc}). The downloaded samples were measured experimentally using the Illumina HiSeq 2000 RNA Sequencing platform by the University of North Carolina TCGA genome characterization center.
First, the genes with missing values were excluded from the data. Genes with low expression levels and those with low interquartile variability between the samples were discarded.  Eventually, the final dataset has been reduced to $p=5747$ genes and $n=606$ samples. 
The gene sets were obtained from Gene Ontology Database (\cite{ashburner2000gene}). To ensure that the GO terms were not too thinly or thickly spread, we extract only those gene sets that contain between 10 and 200 genes. This way the total list reduced to 3295 gene sets, which are used in the analysis. 

\subsection{\textit{E.coli} gene expression data}
We apply the DW-Lasso to a well-known public \textit{E.coli} microarray dataset available in Many Microbe Microarrays database (M3D) (\cite{faith2007many}). The M3D includes microarray  datasets measured using Affymetrix GeneChip genome arrays. 
We select the dataset that contains the expression level of 4297 genes from 466 samples. As a gold standard for performance evaluation, we choose experimentally validated interactions from a curated database, RegulonDB (\cite{salgado2012regulondb}) that comprises a high-confidence level set of interactions supported by genome-wide transcription factor data (\cite{harbison2004transcriptional}). This database contains  the interaction information between transcription factors and genes. To make the list of genes in both microarray and interaction data comparable, we extract 1346 genes that are unique for both datasets. As  a result, our gold standard data contains the list of 2069 interactions between 147 transcription factors and 1199 genes. The final microarray dataset used in the study contains 1346 genes and 466 samples.

\begin{figure*}[!t]%figure1
%		\centerline{\includegraphics[scale=0.25]{Figures/Figure7}}
				\centerline{\includegraphics[scale=0.25]{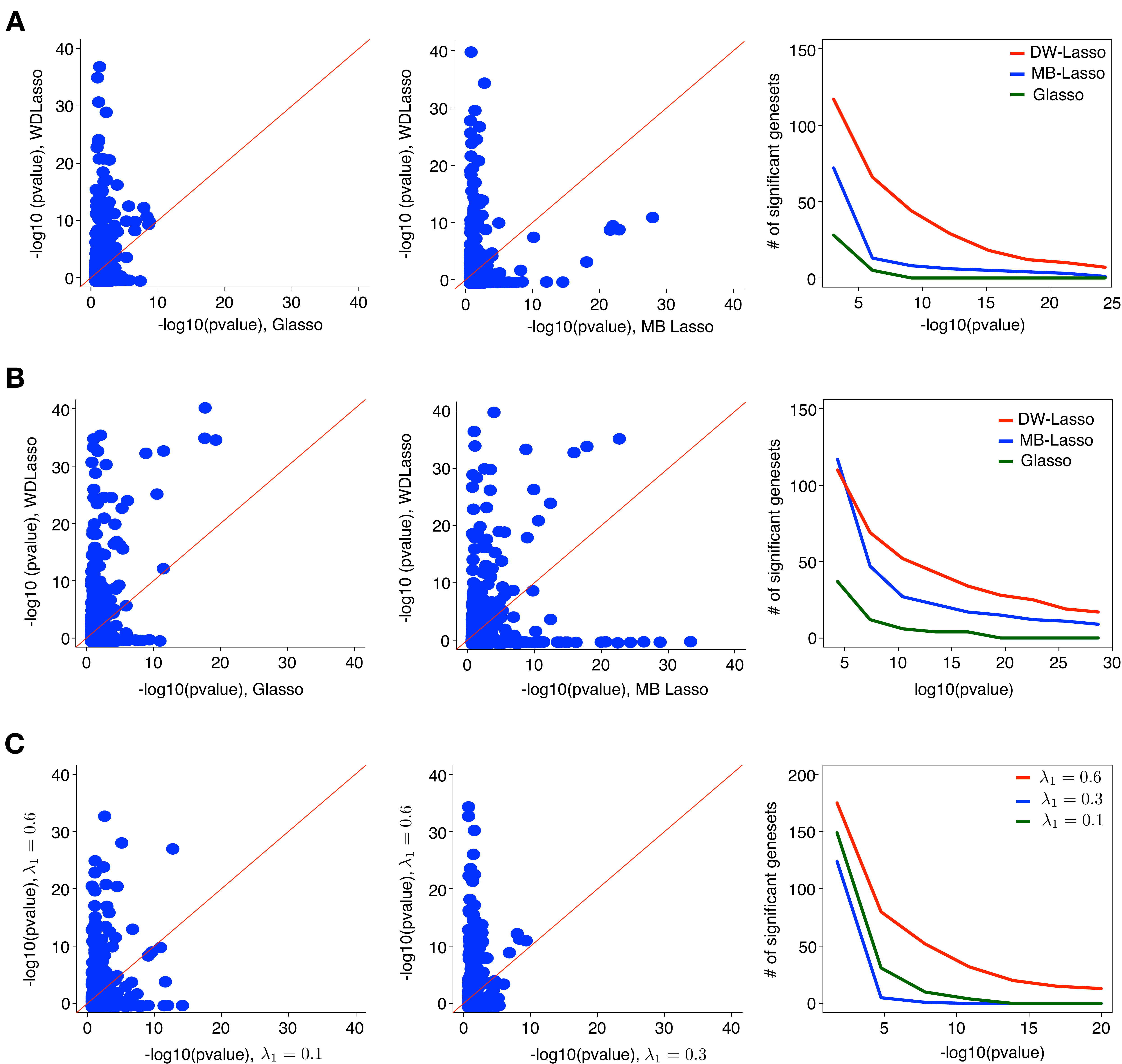}}
	\caption{{ \small Strength of association between gene sets from GO ontology and networks inferred with DW-Lasso, nodewise regression (MB-Lasso) and Glasso from Kidney Clear Cell Carcinoma ((A) compactness score and (B) Knet score). The comparison of strength of association of gene sets for (left) DW-Lasso and Glasso networks and (middle) DW-Lasso and MB-Lasso networks. (right) Quantification of genesets that significantly cluster on inferred networks. (C) (left) and (middle) Strength of association between gene sets from GO ontology and networks inferred with DW-Lasso with different penalty parameters from Kidney Clear Carcinoma  (compactness score). (right) Quantification of genesets that significantly cluster on inferred networks.}}\label{fig:10}
\end{figure*}

\subsection{Quantifying the functional association between GO terms and inferred networks}
Most of publicly available human interaction networks from databases are of low coverage and cannot be used for validation of inferred networks from experimental data. We therefore use an indirect approach and quantify the functional association between publicly available gene sets from Gene Ontology Database and inferred networks.
To quantify this association, we compute two metrics: The compactness score that quantifies the average distance between the genes in the network (\cite{glaab2010extending}) and the Knet score implemented in the SANTA package (\cite{cornish2014santa}) that takes into account the global topology of the network. More information about these metrics can be found in Section \ref{sec:genesets_metrics}. 
We select $\lambda_{1}$ by stability selection and tune $\lambda_{2}$ to infer networks with different sparsities. 
We then tune the penalty parameters of nodewise regression and Glasso such that we get the network with the same sparsity as the network inferred by DW-Lasso. Since the selected metrics are highly sensitive against the missed edges among the nodes in the graph, we tune $\lambda_{2}$ so that the resulting graphs have as few isolated nodes as possible. 
Thus, to allow for a fair comparison, we infer networks that contain $10^5$ interactions for 5747 genes.
The results show that the 3295 gene sets cluster more strongly on DW-Lasso networks than on networks inferred by nodewise regression (MB-Lasso) and Glasso (Figure \ref{fig:10}A (compactness score) and Figure \ref{fig:10}B (Knet score)). This indicates that the DW-Lasso networks are functionally more informative than the networks inferred by nodewise regression (MB-Lasso) and Glasso. In addition, we observe that the clustering of gene sets gets better as $\lambda_{1}$ is chosen such that the number of false positive edges is controlled (Figure \ref{fig:10}C (compactness score)). 
This indicates that the underlying network may contain modular structures that include highly connected genes.
The top GO terms strongly enriched with DW-Lasso networks are related to chromatin remodeling (histone lysine methylation (p-value $ < 1 \times 10^{-24}$), histone H3-K4 methylation (p-value $ < 1 \times 10^{-17}$)), and NF-KB signaling (p-value $ < 1 \times 10^{-10}$). These processes were reported to participate in the development and progression of the kidney cancer (\cite{cancer2013comprehensive}; \cite{ho2016high}). We also observe significant clustering of mitochondrial processes on DW-Lasso networks (mitochondrial respiratory chain (p-value $ < 1 \times 10^{-48}$), mitochondrial respiratory chain complex I (p-value $ < 1 \times 10^{-11}$) and mitochondrial electron transport (p-value $ < 1 \times 10^{-19}$)). It was reported that mitochondrial dysfunction is common in cancer and the mitochondrial electron transport is often affected in carcinogenesis (\cite{ellinger2016systematic}). 
We next check the list of highly ranked genes by degree in the inferred graph. Our analysis reveals several interesting genes that were reported to have a transcription factor activity. For example, TCF4 was demonstrated to play an important role in the progression of kidney cancer via an inhibition of apoptosis (\cite{shiina2003human}). Another highly ranked inhibitor of apoptosis is HSP70 (\cite{gabbert2007expression}).  RUNX1 was reported to be upregulated in various kidney cancers and be potentially used as targets for new therapies (\cite{xiong2014rna}). One study showed the overexpression of ACAT1 in kidney cancer tissues compared to normal tissues (\cite{osunkoya2009diagnostic}). The protein encoded by gene ALDH6A1 was found to be highly expressed in kidney cancer tissues and involved in metabolic processes that are associated with apoptosis and tumorigenesis and thus may play a critical role in renal carcinoma oncogenesis (\cite{perroud2009grade}). Another interesting gene is UQCRC1 which was identified as a potential biomarker. Dysregulation of this gene is involved in impaired mitochondrial electron transport chain function (\cite{ellinger2016systematic}). 

\subsection{Performance of DW-Lasso on \textit{E.coli} data}
\begin{figure}[tp]%figure1
%	\centerline{\includegraphics[scale=0.35,bb=0 0 521 359]{Figures/Figure8}}
		\centerline{\includegraphics[scale=0.35,bb=0 0 521 359]{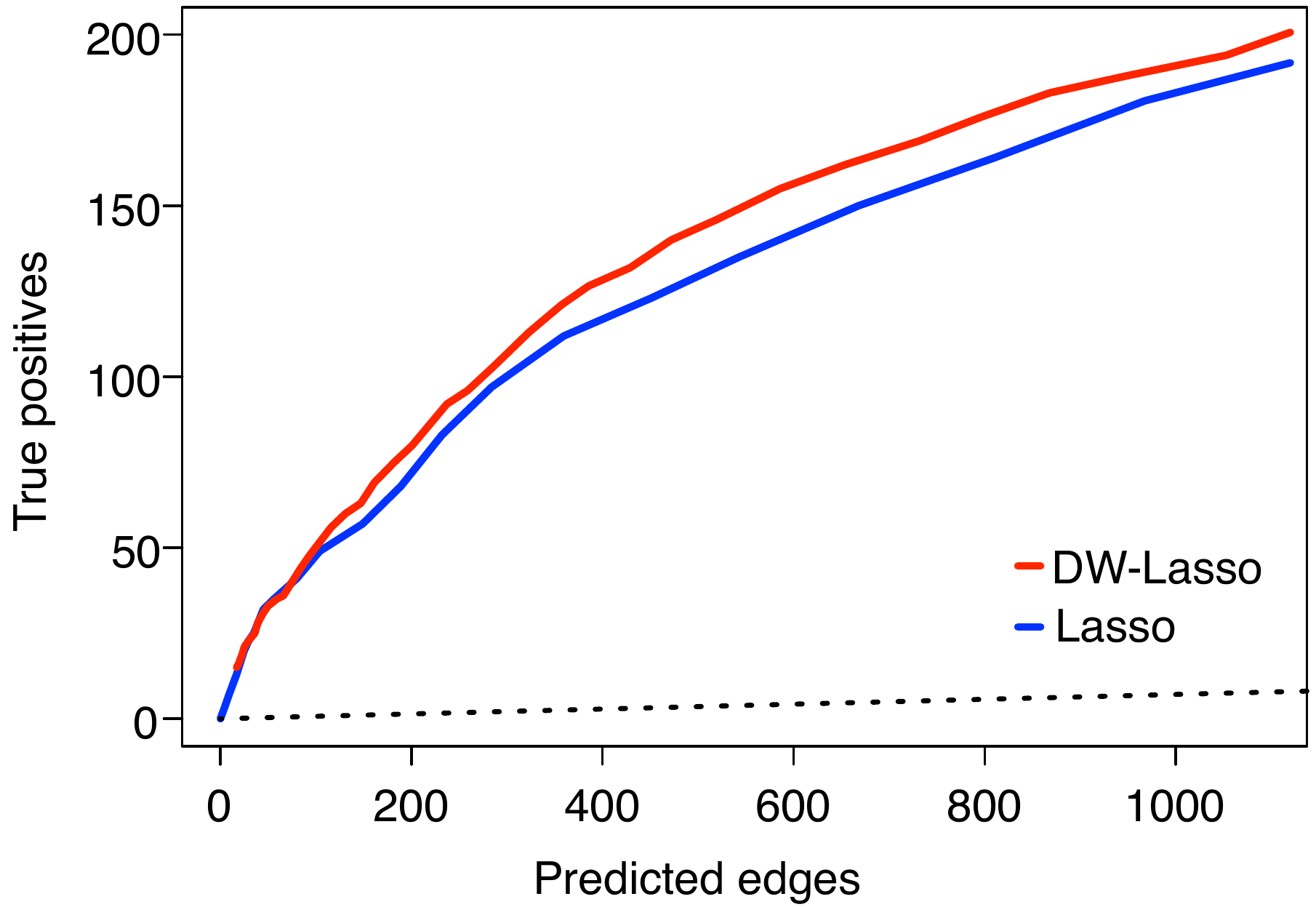}}
	\caption{{ \small Performance of DW-Lasso and Lasso on \textit{E.coli} data. Illustrated is the number of correctly predicted edges vs total predicted edges. The simulation for DW-Lasso is performed with $\lambda_{1}=0.001$ and varying $\lambda_{2}$, whereas for Lasso the penalty parameter is varied.}}\label{fig:11}
\end{figure}

We apply DW-Lasso designed for the inference of transcription factor-gene networks (Section \ref{sec:DW_Lasso_TF}) on \textit{E.coli} data and compare it with the standard Lasso. Since we are interested in reconstructing the network of TF-genes, Glasso and nodewise regression are not relevant in this case. 
We evaluate the performance of methods based on estimated edges and detected true positive edges. For this purpose, the overall sparsity inducing penalty parameters for DW-Lasso and Lasso are varied until both methods detect 1000 edges. 
Results show that both methods are able to detect more edges compared to a random guess (Figure \ref{fig:11}). In addition, DW-Lasso is able to predict relatively more true interactions compared than Lasso. This indicates that the accounting for hub information is advantageous in improving the accuracy of network inference.

\section{Conclusion}
We proposed 'DW-Lasso', a new method  for the inference of hub graphical models from high-dimensional data. 
The proposed method consists of two steps and includes two penalty parameters. Tuning these parameters controls the number of hubs, degree of hubs and overall sparsity of the network.  We consider stability selection criteria (\cite{meinshausen2010stability}) to select the penalty parameters.
With simulation studies considering hub and scale-free graphs, we demonstrate the increased performance of our method in comparison to traditional methods under the small $n$ large $p$ scenario. Additionally, for kidney cancer data, we show good performance of our method by quantifying the functional content of inferred gene networks with GO terms. Clustering results suggest that gene networks might contain modules consisting of genes centered around hub genes. The advantage of our method over the existing methods is that it can naturally infer hub graphs with various hub sparsities, controlled only by a single penalty parameter. 
DW-Lasso performs well on both hub network and not-too-sparse scale-free networks. It relies on less constraining assumptions about the network topology compared to that of concurrent methods, although extreme sparsity cannot be accounted for. 
 Although our method is designed for the analysis of gene expression data, it can also be applied to proteomics, metabolomics and in general to any high dimensional data where a underlying hub structure is expected.

\section*{Software}
The R package is available under the GNU General Public Licence at https://cran.r-project.org/package=DWLasso.\\

\section*{Acknowledgements}
We would like to thank Wasiur Rahman Khuda Bukhsh and Tim Prangemeier for useful comments and discussions. Part of the computations were performed at the Vital-IT Center for high-performance computing (http://www.vital-it.ch) of the SIB (Swiss Institute of Bioinformatics).
\vspace*{-12pt}

\section*{Funding}

This work has been supported by the e:Bio project HostPathX funded by Federal Ministry of Education and Research (BMBF) and Research, Technology and Development Project grant from SystemsX.ch. HK also acknowledges support from the LOEWE research priority program CompuGene and from the H2020 European project PrECISE. \vspace*{-12pt}

\bibliographystyle{plainnat}
\bibliography{document}

\end{document}